\documentclass[aps,prb,onecolumn,groupedaddress,showpacs]{revtex4-1}
\usepackage{graphicx}
\usepackage{latexsym}
\usepackage{fancybox}
\usepackage{color}

\begin{document}
\title{Strong Valence Fluctuation Effects in Sm$Tr_2$Al$_{20}$($Tr=$Ti, V, Cr)}
\author{Akito Sakai and Satoru Nakatsuji}
\affiliation{Institute for Solid State Physics, University of Tokyo, Kashiwa, Chiba 277-8581, Japan}
\date{\today}

\begin{abstract}
We present a single crystal study of low temperature magnetism and transport in Sm$Tr_2$Al$_{20}$ ($Tr =$ Ti, V 
and Cr). Strong valence fluctuation is manifested as Kondo effects including a large Sommerfeld coefficient $\gamma$, a weak 
temperature dependence of magnetic susceptibility and a $-\ln T$ dependent resistivity. 
All the systems order antiferromagnetically at $T_{\rm N}=$ 6.4 K (Ti), 2.9 K (V) and 1.8 K (Cr). 
A systematic change in the susceptibility, specific heat, and resistivity indicates that stronger $c$-$f$ hybridization in SmV$_2$Al$_{20}$ and 
SmCr$_2$Al$_{20}$ than in SmTi$_2$Al$_{20}$ suppresses $T_{\rm N}$ and induces the valence fluctuations, and moreover, field-insensitive heavy fermion states.
\end{abstract}


\maketitle
Sm-based intermetallics have attracted much attention in the research of strongly correlated 
$f$-electron systems for their rich variety of physical phenomena.
One significant characteristic property of the Sm ion is the valence degree of freedom.
For instance, Sm$X$($X$=Te, Se, S) exhibits a semiconductor-metal transition under pressure, showing a jump in valence\cite{SmX1,SmX2}. 
SmB$_6$ is known as a canonical Kondo insulator with intermediate valence\cite{SmB6ins}. Although quite rare in Sm-based systems, the Kondo effect is observed in SmSn$_3$ and SmFe$_4$P$_{12}$ as $-\ln T$ dependence of the resistivity\cite{SmSn3,SmFe4P12}.
Recent extensive work on filled skutterudite compounds \cite{Maple2007182,Sato2007188}has 
elucidated that when located at a highly symmetric center of a cage, the Sm ion may induce interesting physical properties. 
SmRu$_4$P$_{12}$ shows a metal-insulator transition at $T_{\rm MI}=16.5$ K and 
an octupole order below $T_{\rm MI}$ has been proposed to explain the sharp drop in the elastic constant\cite{Sekine2000,SmRu4P12oct}.
Unusual field-insensitive heavy-fermion
behavior has been discovered in SmOs$_4$Sb$_{12}$\cite{Sanada2005,Yuhasz2005}. Several candidates of the mechanism are proposed, such as 
mixed valence\cite{Mizumaki2007,Yamasaki2007,Kotegawa2007,Tsubota2008} and anharmonic oscillation in a cage called the rattling vibration\cite{Matsuhira2007,Ogita2007,Hattori2005,Hotta2007}.

The CeCr$_2$Al$_{20}$-type compounds $RTr_2$Al$_{20}$ ($R$: rare earth; $Tr$: transition metal ) 
provide another cage structure similar to the filled skutterudite compounds. 
They have a space group $Fd{\bar 3}m$ and the symmetry of the $R$ site is $T_{\rm d}$ and cubic. The number of Al constructing the cage is 16, in comparison with 12 for the filled skutterudites and is the largest for the coordination
 number of tetrahedral groupings of spheres\cite{FrankKasper}.
It suggests that the number of channels of Kondo coupling should be large in this system, and in fact, 
our recent study on the Pr$Tr_2$Al$_{20}$ systems reveals strong hybridization effects between Pr-4$f$ electrons and conduction($c$) electrons \cite{Akito}. 
Therefore, the Sm analogs may exhibit strong $c$-$f$ hybridization effects  and thus show interesting phenomena.

In this Rapid Communication, we report the results of single-crystal studies on the low-temperature magnetism and transport of Sm$Tr_2$Al$_{20}$ ($Tr$= Ti, V, Cr).
Antiferromagnetic orders are observed at $T_{\rm N}=$6.4 K (Ti), 
2.9 K (V) and 1.8 K (Cr). The entropy of these compounds saturates to $R\ln4$, showing that the magnetic order is based on the $\Gamma_8$ quartet of the $J=5/2$ multiplet.
Interestingly, the enhancement of the Sommerfeld coefficient $\gamma$ is observed as $T_{\rm N}$ is suppressed by substitution, which reaches 1000 (mJ/mol K$^2$) for SmCr$_2$Al$_{20}$.
Magnetic susceptibilities exhibit nearly temperature independent behavior, indicating strong valence fluctuation.
The systematic change in the susceptibility and the $-\ln T$ dependence of the resistivity reveals 
that the $c$-$f$ hybridization becomes enhanced in the sequence from $Tr =$ Ti, V, to Cr, inducing the Kondo effect, which has been rarely seen in Sm-based Kondo lattice systems. Moreover, our specific heat measurements found that the low-temperature heavy fermion state is significantly insensitive to magnetic field at least up to 9 T, suggesting a role of the nonmagnetic Sm$^{2+}$ state stabilized by the valence fluctuation.

Single crystals of Sm$Tr_2$Al$_{20}$ ($Tr$= Ti, V, Cr) were grown by the Al self-flux method under vacuum, using 4$N$(99.99\%)-Sm, 3$N$-$Tr$ and 5$N$-Al.
The starting ratio of the elements was 1 : 2 : 45. They were prepared in alumina crucibles and
sealed in silica tubes and slowly cooled from $1150$ to $750\ {}^\circ\mathrm{C}$ for 60 h.
The crystal structure was verified to be of cubic CeCr$_2$Al$_{20}$ type with the space group $Fd{\bar 3}m$ by the x-ray powder and single-crystal diffraction measurements and is found to have lattice parameters $a$ = 14.723 \AA (Ti), 14.571 \AA (V), 14.501 \AA (Cr), all of which are consistent with the previous report \cite{chemistry}.  
The electrical resistivity $\rho$ and specific heat $C_P$ above 0.4 K were measured by the standard four-probe dc method and a thermal relaxation method, respectively.
The dc magnetic susceptibility $\chi$ above 2 K  was measured by a commercial superconducting quantum interference device (SQUID) magnetometer.
\begin{figure}[t]
\begin{center}
\includegraphics[keepaspectratio, scale=1.3]{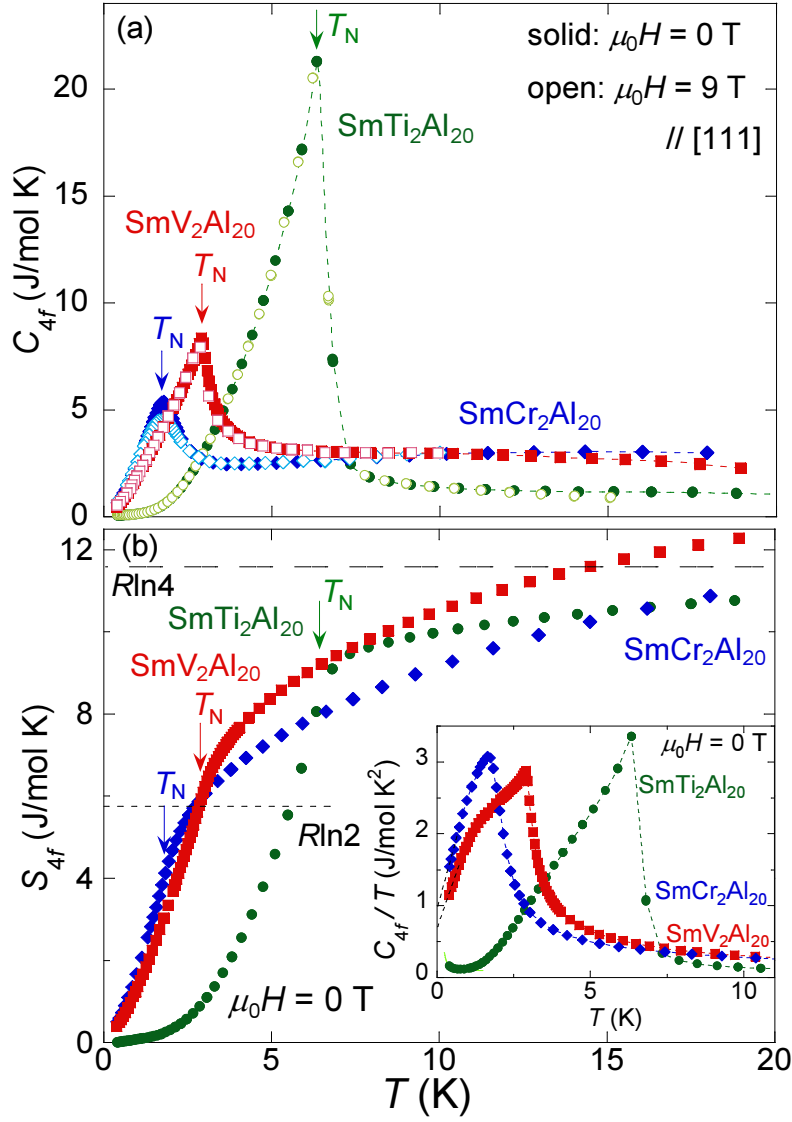}
\caption{(Color online) (a) Temperature dependence of the $4f$ contribution 
to the specific heat, $C_{4f}$, for SmTi$_2$Al$_{20}$ (circle), SmV$_2$Al$_{20}$ (square) and  SmCr$_2$Al$_{20}$ (diamond) under $\mu_0 H = 0$ T (solid symbols) and 9 T (open symbols). 
Arrows indicate the transition temperatures defined as the peak temperature of $C_P(T)$
(b) Temperature dependence of the $4f$ electron contribution $S_{4f}$ to the entropy. The horizontal dashed lines indicate
$S_{4f} = R\ln2$ and $R\ln4$, respectively. Inset: $T$ dependence of $C_{4f}/T$. 
To estimate the Sommerfeld coefficient $\gamma$ at $T = 0$ and $S_{4f}$ at 0.4 K, a linear extrapolation (dashed straight line) of $C_{4f}/T$ to $T = 0$  is used.
The solid line indicates the fitting to the form $C/T=\gamma +A/T^3$ for SmTi$_2$Al$_{20}$}\label{Cp}
\end{center}
\end{figure}

Figure \ref{Cp}(a) shows the 4$f$ electron contribution to the specific heat 
$C_{4f}$ under fields of 0 T (solid symbols) and 9 T (open symbols).
This was estimated by subtracting $C_P$ of the La analogs, namely La$Tr_2$Al$_{20}$ \cite{Akito}, from $C_P$ of Sm$Tr_2$Al$_{20}$ ($Tr = $Ti, V, Cr).
A sharp lambda-like anomaly of $C_{\mathrm{4}f}$ is observed at $T_{\rm N} = 6.4$ K for SmTi$_2$Al$_{20}$, $T_{\rm N} = 2.9$ K for SmV$_2$Al$_{20}$
 and $T_{\rm N} = 1.8$ K for SmCr$_2$Al$_{20}$, indicating a second-order phase transition.
 The temperature dependence of the entropy $S_{\mathrm{4}f}$ estimated by integrating $C_{4f}/T$ vs. $T$ is shown in Fig.1 (b).
$S_{\mathrm{4}f}(0.4$ K) and the Sommerfeld coefficient $\gamma$ at $T = 0$ were estimated assuming a linear decrease of $C_{4f}/T$ between $T_{\rm N}$ and $T = 0$ for SmV$_2$Al$_{20}$ and SmCr$_2$Al$_{20}$ (Fig. \ref{Cp}(b) broken lines).
As for SmTi$_2$Al$_{20}$, the upturn of $C/T$ below 1 K is ascribable a nuclear Schottky component, and we used the fitting form $C/T=\gamma +A/T^3$ (Fig. \ref{Cp}(b) inset, solid line).
$S_{\mathrm{4}f}(T)$ for all systems shows a saturation to $\sim R\ln4$ at $T \sim 20$ K, indicating the ground quartet (Fig. \ref{Cp}(b)).
If we assume that the 4$f$ electrons of Sm$^{3+}$ are localized, $J=5/2$ multiplet splits into the  doublet $\Gamma_7$ and the  
quartet $\Gamma_8$ because of the $T_d$ cubic symmetry at the Sm site. Thus, the crystalline electric 
field (CEF) ground state should be the $\Gamma_8$ quartet and the gap $\Delta$ between the two CEF states should be larger than 40 K.
The large Sommerfeld coefficient $\gamma$ of $\sim$ 100 (mJ/mol K$^2$) for SmTi$_2$Al$_{20}$, $\sim$ 720 (mJ/mol K$^2$) for SmV$_2$Al$_{20}$ and $\sim$ 1000 (mJ/mol K$^2$) for SmCr$_2$Al$_{20}$
 below $T_{\rm N}$ suggests that the electron mass is enhanced by a many-body correlation effect.
In this type of Kondo lattice systems, the origin of the mass enhancement must be the $c$-$f$ hybridization, and thus the increase in $\gamma$ points to the corresponding increase in the hybridization strength
and the approach toward a quantum critical point from a magnetic ordered side.
Surprisingly, the shape of the specific heat including the large Sommerfeld coefficient $\gamma$ does not change even under a field of 9 T (Fig1. (a)). 
It indicates that the ordered phase is insensitive to the magnetic field.
A similar field insensitivity is sometimes observed in Sm compounds such as SmSn$_3$, SmOs$_4$P$_{12}$ and SmPt$_4$Ge$_{12}$\cite{SmSn3,SmOs4P12,SmPt4Ge12}.

\begin{figure}[t]
\begin{center}
\includegraphics[keepaspectratio, scale=1.5]{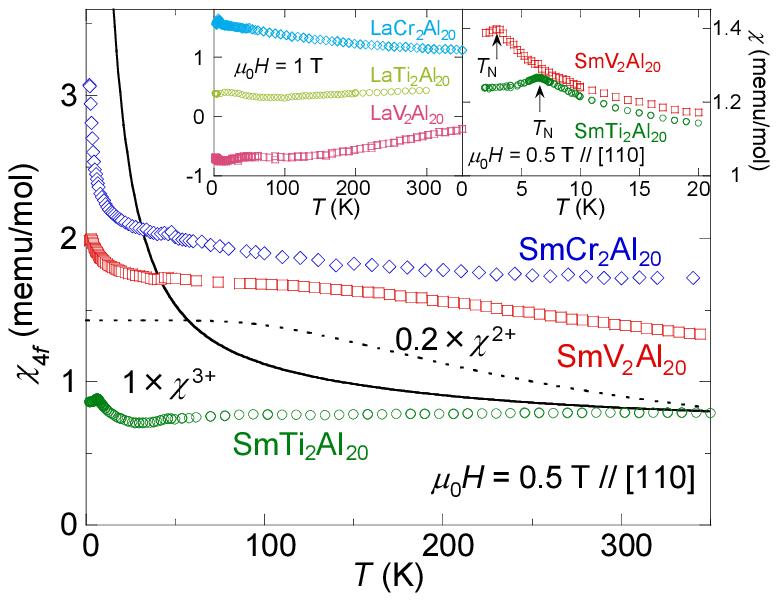}
\caption{(Color online) Temperature dependence of the magnetic susceptibility of the $4f$ electron contribution $\chi_{4f}$ for SmTi$_2$Al$_{20}$ (circle),  
SmV$_2$Al$_{20}$ (square) and  SmCr$_2$Al$_{20}$ (diamond), and under $\mu_0 H =  0.5$ T $\parallel [110]$. No anisotropy is found under the field below 0.5 T. 
The solid and dotted lines show $\chi^{3+}(T)$ and $\chi^{2+}(T)$ calculated for $4f^{5}$ $J=5/2$ and $4f^{6}$ $J=0$ single ion ground states, respectively.
Left-hand inset: $\chi(T)$ of La$Tr_2$Al$_{20}$. 
Right-hand inset: Low temperature part of $\chi$ for SmTi$_2$Al$_{20}$ and SmV$_2$Al$_{20}$. The arrows indicate
 the transition temperatures defined as the cusp temperatures in $\chi(T)$.}\label{chi}
\end{center}
\end{figure}

Figure \ref{chi} shows the temperature dependence of the 4$f$ electron contribution to the magnetic susceptibility $\chi_{4f}$.
Here, $\chi_{4f}$ is estimated as the difference between $M/H$ of Sm$Tr_2$Al$_{20}$ and La$Tr_2$Al$_{20}$ 
and is found isotropic under $\mu_0 H \leq 0.5$ T and above $T_{\rm N}$. 
$\chi(T)$ of La$Tr_2$Al$_{20}$ is found nearly $T$-independent and Pauli paramagnetic (Fig \ref{chi} inset).
Solid and dotted lines show $\chi^{3+}(T)$ and $\chi^{2+}(T)$ calculated for Sm$^{3+}$ $4f^{5}$ $J=5/2$ and Sm$^{2+}$ $4f^{6}$ $J=0$ single ion ground states, respectively. 
Here, we adopt a typical value of the spin orbit interaction $\lambda/k_{\rm B} = 580$ K (Sm$^{3+}$), 420 K (Sm$^{2+}$)  \cite{So.SmB6}, 
CEF parameters $A_4\langle{r^4}\rangle/k_{\rm B}=-500$ K and $A_6\langle{r^6}\rangle/k_{\rm B}=150$ K, leading to the CEF gap of $\sim200$ K between $\Gamma_8$ and $\Gamma_7$\cite{Sm.CEF}. 

Clearly, $\chi_{4f}$ of Sm$Tr_2$Al$_{20}$ shows a weaker temperature dependence than the calculation for the $\chi^{3+}(T)$. 
Only the upturn below $\sim 20$ K may come from the 
$J=5/2$ multiplet. On the other hand, the temperature independent susceptibility at 50 K $<T<$ 150 K for SmV$_2$Al$_{20}$ is ascribable to the Van Vleck susceptibility due to the off-diagonal term between $J=0$ ground 
state and $J=1$ excited state of Sm$^{2+}$ ion. 
In addition, $\chi_{4f}$ of the V and Cr compounds at 350 K is larger than the results of Sm$^{3+}$ CEF calculation. It indicates the existence of Sm$^{2+}$ component, 
often observed in valence fluctuating Sm compounds such as SmS under pressure and SmB$_6$\cite{Maple,Cohen}.
By using the susceptibility at 350 K, we roughly estimate the ratio between Sm$^{3+}$ and Sm$^{2+}$ 
to be 0.8 : 0.2 for SmV$_2$Al$_{20}$ and  0.7 : 0.3 for SmCr$_2$Al$_{20}$. 
However, the experimental data of  $\chi_{4f}$ is not fully reproduced by the sum of divalent and trivalent component of Sm ion, namely, $w\chi^{3+}(T)$+$(1-w)\chi^{2+}(T)$ below 200 K with the weight $w$ for the Sm$^{3+}$ component. This temperature scale is too high for antiferromagnetic fluctuations to be effective in this three dimensional material with low $T_{\rm N}$. Instead, it indicates that the ratio between Sm$^{3+}$ and Sm$^{2+}$ is possibly temperature dependent and the valence decreases on cooling as found in SmOs$_4$Sb$_{12}$\cite{Mizumaki2007,Yamasaki2007}. 
Notably, the fact that the value of $\chi_{4f}$ at high temperatures 
is enhanced by substitution, corresponding to the increase in the Sm$^{2+}$ component, indicates the valence fluctuations increase in order from SmTi$_2$Al$_{20}$, SmV$_2$Al$_{20}$, to SmCr$_2$Al$_{20}$.

The inset to Fig. \ref{chi} shows the low temperature part of the magnetic susceptibility $\chi$ for SmTi$_2$Al$_{20}$ and SmV$_2$Al$_{20}$ near $T_{\rm N}$.  
The arrows show the transition temperature defined at the cusp of $\chi$, which is almost consistent with the peak in the specific heat (Fig. \ref{Cp}(a)).
$T_{\rm N}$ for SmCr$_2$Al$_{20}$ is lower than 2 K, the lowest $T$ of the measurement.
The decrease of $\chi$ below $T_{\rm N}$ indicates that the phase transitions seen in SmTi$_2$Al$_{20}$ and SmV$_2$Al$_{20}$ are 
antiferromagnetic transitions. The muon spin relaxation measurements have been carried out for SmTi$_2$Al$_{20}$ and SmCr$_2$Al$_{20}$, and showed the development of the internal field below $T_{\rm N}$ \cite{muSR}, consistent with the antiferromagnetism. 

\begin{figure}[t]
\begin{center}
\includegraphics[keepaspectratio, scale=1.3]{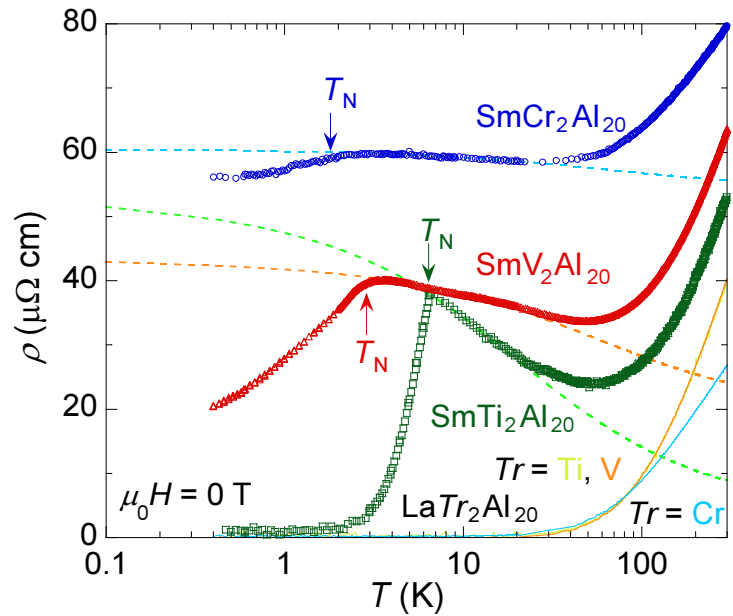}
\caption{(Color online) Temperature dependence of the electrical resistivity $\rho$ of Sm$Tr_2$Al$_{20}$ ($Tr=$Ti, V, Cr). The arrows indicate
 the transition temperatures defined at the peak temperature of the specific heat. 
The dashed lines indicate the fitted curve using Hamann's expression\cite{Hamann} (see text).The solid lines indicate the inelastic component of the resistivity
 $\rho(T)-\rho_{0}$ of La$Tr_2$Al$_{20}$ with a residual resistivity $\rho_{0} = 0.5$ $(\mu\Omega$cm) (Ti), 11.6 $(\mu\Omega$cm) (V), and 51 $(\mu\Omega$cm) (Cr).}\label{rho}
\end{center}
\end{figure}

Figure \ref{rho} shows the temperature dependence of the electrical resistivity $\rho(T)$ of the Sm$Tr_2$Al$_{20}$ systems.
All Sm compounds exhibit a resistivity drop due to the magnetic order at $T_{\rm N}$.
$T_{\rm N}$ defined as the peak temperature of the specific heat are indicated by arrows and are consistent with the cusp in $\rho(T)$.
Interestingly, $\rho(T)$ shows a $-\ln T$ dependence at $T_{\mathrm N}<T<30 \sim 40$ K.
This is reminiscent of the Kondo effects seen in Ce, Yb, and U-based heavy-fermion compounds \cite{HewsonBook}.
The CEF scheme alone cannot produce the $-\ln T$ dependence but only the $T$ independent $\rho$ in the $T$ regime between $T_{\rm N}$ and the CEF gap $\Delta$, which should be higher than 40 K for the current systems. 
In addition, since the $\rho (T)$ of La$Tr_2$Al$_{20}$ exhibits almost no temperature dependence in this $T$ region,
a phonon mechanism can be ruled out (Fig. \ref{rho}).
Note that the critical effects associated with the phase transition 
cannot be the origin of the increase of $\rho(T)$ starting below $T\sim$40 K$\gg T_{\rm N}$.
Thus, it may well come from the Kondo effect based on the hybridization between conduction electrons and the 
 magnetic $\Gamma_8$ state.
In Sm compounds, such a Kondo effect has been rarely found, so faronly in SmSn$_3$\cite{SmSn3} and SmFe$_4$P$_{12}$\cite{SmFe4P12}, to the best of our knowledge.

\begin{figure}[t]
\begin{center}
\includegraphics[keepaspectratio, scale=1]{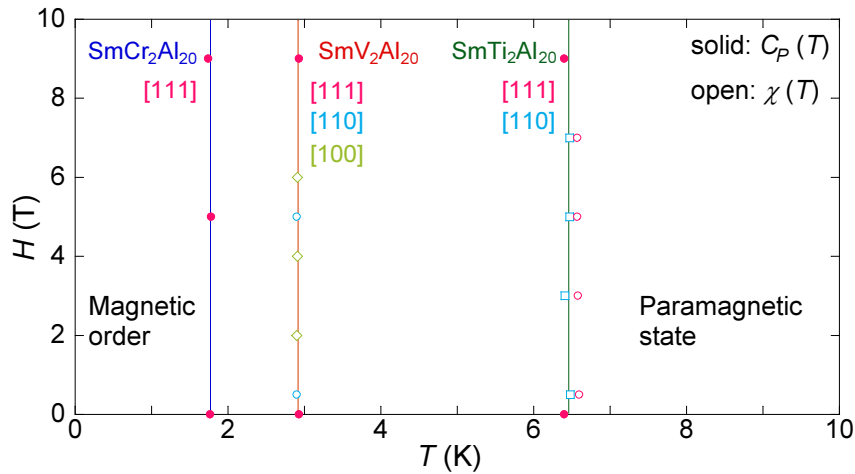}
\caption{(Color online) $T$-$H$ phase diagram of Sm$Tr_2$Al$_{20}$ ($Tr=$Ti, V, Cr) constructed by measurements made under fields along [100] (diamond), [110] (square) and [111] (circle). 
The open symbols are determined from the cusp of $\chi(T)$
 and solid symbols are determined from the peak of $C_P(T)$. The solid lines are guides to the eye.}\label{Phase}
\end{center}
\end{figure}

The $T$-$H$ phase diagram of the Sm$Tr_2$Al$_{20}$ systems constructed by measurements under fields along $[100]$, $[110]$ and $[111]$
 is shown in Fig. \ref{Phase}. For all compounds, no field direction dependence is seen, indicating the isotropic nature of the phase boundaries.
The transition temperature $T_{\rm N}(H)$ is determined as the peak $T$ 
of $C_P(T)$ (solid symbols) and $\chi(T)$ (open symbols).
In each system, the $T_{\rm N}(H)$ boundary line is almost vertical to the temperature axis, reflecting the field insensitivity of the transition temperature.

Now let us discuss the chemical trend in the valence fluctuations. 
According to Hamann's expression on the resistivity\cite{Hamann}, the smaller slope in the logarithmic dependence of $\rho(T)$ 
indicates a larger Kondo temperature $T_{\rm K}$. 
In fact, $\rho(T)$ in Fig. \ref{rho} can be fit to the expression (dased lines) with $T_{\rm K} \sim 20\pm$5 K ($Tr =$ Ti), $55\pm$5 K (V), and $61\pm$20 K (Cr) for $S=1/2$. 
Although the absolute value may not be correct because the model does not take into account the quartet CEF ground state,
the result at least captures the correct trend, namely, the strongest Kondo effect
in the Cr and the weakest in the Ti compound. 
Generally, the Kondo effect suppresses the entropy release due to a magnetic transition. In fact,
the entropy $S_{4f}$ reaches $\sim$ 70 $\%$ ($Tr =$ Ti), 45 $\%$ (V) and 35 $\%$ (Cr) of $R\ln 4$ at $T_{\rm N}$ (Fig. \ref{Cp}(b)).
Besides, SmCr$_2$Al$_{20}$ has the largest $\chi$ at 300 K, while SmTi$_2$Al$_{20}$ has the smallest.
All these confirm that SmCr$_2$Al$_{20}$ has the strongest valence fluctuations, and SmTi$_2$Al$_{20}$ has the weakest.

This systematic increase in the Kondo temperature and valence fluctuation scale suggests the corresponding enhancement in the $c$-$f$ hybridization in order from $Tr =$ Ti to Cr.
The replacement of Ti by V and Cr compresses the unit-cell volume by $\sim$ 3 and 5 \% respectively. For Sm based intermetallics, the (chemical) pressure normally decreases the Kondo temperature and increases the magnetic ordering temperature\cite{Barla2004,Kotegawa2005}. 
Therefore, additional one (two) $3d$ electron(s) of Cr(V) in comparison with Ti likely play an important role in enhancing the $c$-$f$ hybridization and inducing the valence fluctuations.

Furthermore, the strong valence fluctuations may be the origin for the field insensitivity of the large Sommerfeld coefficient $\gamma$ and the magnetic phase boundary. The $g$ factor for Sm compounds is small and thus significantly reduces the Zeeman energy scale of the CEF ground state in comparison with, for instance, Ce compounds\cite{SmGFactor}. However, the smallness of the $g$ factor alone cannot explain the robust feature against such a large field as 9 T.
Instead, the nonmagnetic Sm$^{2+}$ state stabilized by the valence fluctuations might play an important role in making the ground state robust against the magnetic field.

A similar type of the remarkably stable heavy fermion state has been discovered in SmOs$_4$Sb$_{12}$\cite{Sanada2005} and has attracted much attention. 
To explain such a magnetically robust heavy fermion state, the rattling of the $f$-electron has been inferred as a plausible mechanism \cite{Matsuhira2007,Ogita2007,Hattori2005,Hotta2007}. However, 
in our systems, the rattling mechanism is unlikely because no particularly large thermal displacement factor has been found for the Sm ion among the series of $RTr_2$Al$_{20}$\cite{Julia}.

The fact that the increase in the Kondo coupling
and valence fluctuation in turn decreases the ordering temperature 
suggests a proximity to a quantum critical point. 
We would note that,to the best of our knowledge, there has been no systematic study on quantum criticality in mixed valent Sm compounds.
If $T_{\rm N}$ is much suppressed,
 anomalous metallic behavior and superconductivity might emerge in these systems.
Further investigation is necessary to elucidate such possibilities.

We thank T. Sakakibara, T. Ito, W. Higemoto and Y. Nakanishi for useful discussions.
This work is partially supported by Grants-in-Aid (No.21684019) from 
JSPS, by Grants-in-Aids for Scientific Research on Innovative Areas ``Heavy Electrons" of MEXT, Japan, 
and by Grant-in-Aid for JSPS Fellows. 

\bibliographystyle{apsrev4-1}
%

\end{document}